\documentclass[twocolumn,epjc3]{svjour3}
\usepackage{lmodern}
\usepackage{amsmath}
\usepackage{amssymb}
\usepackage{epstopdf}
\usepackage{widetext}

\RequirePackage{fix-cm}
\smartqed  
\RequirePackage{graphicx}
\newcommand{\be}{\begin{equation}}
\newcommand{\ee}{\end{equation}}
\newcommand{\bn}{\begin{eqnarray}}
\newcommand{\en}{\end{eqnarray}}
\newcommand{\bes}{\begin{subequations}}
\newcommand{\ees}{\end{subequations}}

\journalname{Eur. Phys. J. C}

\begin{document}

\title{The Starobinsky model within the $f(R,T)$ formalism as a cosmological model}
\author{P.H.R.S. Moraes$^{1,a}$, R.A.C. Correa$^{1,2,b}$, G. Ribeiro$^{2,c}$}
\date{Received: date / Accepted: date}
\maketitle

\begin{abstract}
In this paper we derive a cosmological model from the $f(R,T)$ theory of
gravity, for which $R$ is the Ricci scalar and $T$ is the trace of the
energy-momentum tensor. We consider $f(R,T)=f(R)+f(T)$, with $f(R)$ being
the Starobinksy model $R+\alpha R^{2}$ and $f(T)=\gamma T$, with $\alpha$
and $\gamma$ constants. We find that from such a functional form, it is
possible to describe the cosmological scenario of a radiation-dominated
universe, which has shown to be a non-trivial feature within the $f(R,T)$ formalism. 

\ 

\end{abstract}




\thankstext{1}{e-mail: moraes.phrs@gmail.com} 
\thankstext{2}{e-mail:
rafael.couceiro@ufabc.edu.br} 
\thankstext{3}{e-mail:
ribeiro.gabriel.fis@hotmail.com} 

\institute{ITA - Instituto Tecnol\'ogico de Aeron\'autica - Departamento de F\'isica, 12228-900, S\~ao Jos\'e dos Campos, S\~ao Paulo, Brazil\\
$^{2}$FEG - Faculdade de Engenharia de Guaratinguet\'a - Departamento de F\'isica e Qu\'imica, 12516-410, Guaratinguet\'a, S\~ao Paulo, Brazil}


\section{Introduction}

\label{sec:int}

The $f(R)$ theories of gravity \cite%
{capozziello/2005,nojiri/2006,nojiri/2008} are an optimistic alternative to
the shortcomings General Relativity (GR) faces as the underlying
gravitational theory for a cosmological model \cite%
{sola/2015,padmanabhan/2003,jamil/2009}. They can account for the cosmic
acceleration \cite{riess/1998,perlmutter/1999}, providing a great match
between theory and cosmological observations \cite%
{amarzguioui/2006,tsujikawa/2008,fay/2007}, and also for inflation \cite%
{rinaldi/2014,huang/2014,myrzakulov/2015,elizalde/2010,myrzakul/2015,barvinsky/2015,bamba/2015}
and dark matter issues \cite{capozziello/2007,lubini/2011,shojai/2014}.

One of the crucial troubles surrounding GR is that apparently it cannot be
quantized, although attempts to do so have been proposed, as String Theory 
\cite{fradkin/1985,witten/1986,friedan/1986} (check also \cite%
{barbon/2004,brink/2004} for reviews on the topic), and can, in future,
provide us a robust and trustworthy model of gravity - quantum mechanics
unification.

Meanwhile it is worthwhile to attempt to consider the presence of quantum
effects in gravitational theories. Those effects can rise from the
consideration of terms proportional to the trace of the energy-momentum
tensor $T$ in the gravitational part of the $f(R)$ action, yielding the $%
f(R,T)$ gravity theories \cite{harko/2011}. Those theories were also
motivated by the fact that although $f(R)$ gravity is well behaved in
cosmological scales, the Solar System regime seems to rule out most of the $%
f(R)$ models proposed so far \cite%
{erickcek/2006,chiba/2007,capozziello/2007b,olmo/2007}.

Despite its recent elaboration, $f(R,T)$ gravity has already been applied to
a number of areas, such as Cosmology \cite%
{mc/2016,ms/2016,mrc/2016,moraes/2015,moraes/2014,moraes/2016,jamil/2012,singh/2014,farasat_shamir/2015,sahoo/2014,rudra/2015,sharif/2013,reddy/2014,sahoo/2016,houndjo/2014}
and Astrophysics \cite%
{mmm/2016,mam/2016,amam/2016,zubair/2015,noureen/2015,noureen/2015b}.

By deeply investigating the outcomes and features of an $f(R,T)$ or $f(R)$
model, one realizes the strong relation they have with the functional form
of the chosen functions and free parameter values. In fact, a reliable
method to constraint those ``free" parameters to values that yield realistic
models can bee seen in \cite{cm/2016} and \cite{cmsdr/2015} for $f(R,T)$ and 
$f(R)$ models, respectively.

In $f(R)$ gravity a reliable and reputed functional form was proposed by
A.A. Starobinsky as \cite{starobinsky/2007}

\begin{equation}  \label{i1}
f(R)=R+\alpha R^{2},
\end{equation}
which is known as Starobinsky Model (SM), with $\alpha$ a constant. It
predicts quadratic corrections of the Ricci scalar to be inserted in the
gravitational part of the Einstein-Hilbert action. 

An analysis of matter density perturbations in SM was presented in \cite%
{fu/2010}. Black hole studies were made for $R^{2}$ gravity in \cite%
{hendi/2011}. The consideration of wormholes in such theories can be
appreciated in \cite{pavlovic/2015,kaneda/2016}.

Our proposal in this paper is to construct a cosmological scenario from an $%
f(R,T)$ functional form whose $R-$dependence is the same as in the SM, i.e.,
with a quadratic extra contribution of $R$, as in Eq.(\ref{i1}). The $T-$%
dependence will be considered to be linear, as $\gamma T$, with $\gamma$ a
constant. Therefore, we will take

\begin{equation}  \label{i2}
f(R,T)=R+\alpha R^{2}+\gamma T.
\end{equation}

Despite the high number of considerations of the SM in $f(R)$ cosmology
(check also \cite{goswami/2013,nojiri/2015,abebe/2016}), it has not been
considered for the $R-$dependence in $f(R,T)$ models for cosmological
purposes so far, only in the study of astrophysical compact objects \cite%
{zubair/2015,noureen/2015,noureen/2015b}. We believe this is due to the
expected high nonlinearity of the resulting differential equation for the
scale factor. Anyhow, the consideration of quantum corrections together with
quadratic geometrical terms can imply interesting outcomes in a cosmological
perspective as it did in the astrophysical level (check \cite%
{zubair/2015,noureen/2015,noureen/2015b}). Therefore we present here a
reliable and well referenced method to obtain solutions for such a
cosmological scenario.

Here let us stress that the $f(R,T)$ formalism exhibits a sort of
shortcoming for a specific era of the Universe evolution. One could ask what
are the predictions of $f(R,T)$ gravity in the regime $T=0$. It is natural
to think that for different functional forms formulated to the $f(R,T)$
function, the regime $T=0$ makes $f(R,T)$ gravity to recover $f(R)$
theories. The regime $T=0$ is achieved for $p=\rho/3$, with $p$ and $\rho$
being the pressure and density of the Universe, respectively, which is the
equation of state (EoS) of radiation. Therefore, from a cosmological
perspective it becomes intuitive to think that $f(R,T)$ gravity itself is
not able to describe the era in which the Universe was dominated by radiation%
\footnote{%
In \cite{ms/2016} it was deeply discussed that this non-contribution regime
of $f(R,T)$ gravity can also be expected in vacuum; for instance, in the
study of gravitational waves propagation.}. It would only recover the $f(R)$ outcomes.

The $T=0$ issue surrounding the $f(R,T)$ formalism was already investigated
in \cite{ms/2016,moraes/2014,moraes/2016,baffou/2015,sun/2016}. In \cite%
{ms/2016}, in order to be able to describe the radiation era of the
Universe, a scalar field was invoked in $f(R,T)$ gravity, namely the $%
f(R,T^{\phi})$ gravity. In \cite{moraes/2014} such a description became
possible only in a five-dimensional space-time, while in \cite{moraes/2016}
the speed of light was considered a variable and an alternative scenario to
inflation was obtained. Here, instead, one of our goals is to check if
restrictively the choice of the SM for the $R$ dependence in the $f(R,T)$
function is able to make $f(R,T)$ formalism to describe a
radiation-dominated universe.

The SM in $f(R)$ formalism is known to successfully describe the accelerated
periods of the Universe evolution, namely the inflationary and dark energy
eras \cite%
{starobinsky/2007,kaneda/2016,sebastiani/2015,farakos/2013,ellis/2013,appleby/2010}%
. Would it also be a powerful tool to help $f(R,T)$ gravity to be able to
describe the radiation era of the Universe? Let us address this question in
the next sections.

\section{An overview of the $f(R,T)$ formalism}

\label{sec:frt}

Originally proposed as a generalization of the $f(R)$ theories, the $f(R,T)$
gravity considers the gravitational part of the model action to be dependent
not only on a general function of the Ricci scalar $R$, but also on a
general function of the trace of the energy-momentum tensor $T$, as

\begin{equation}  \label{frt1}
S_{grav}=\frac{1}{16\pi}\int d^{4}x\sqrt{-g}f(R,T),
\end{equation}
with $g$ being the determinant of the metric and $f(R,T)$ the function of $R$
and $T$. Moreover, throughout this article we will consider natural units.

By varying action (\ref{frt1}) with respect to the metric $g_{\mu\nu}$, one
obtains the following field equations:

\begin{align}
& f_{R}(R,T)R_{\mu \nu }-\frac{1}{2}f(R,T)g_{\mu \nu }+(g_{\mu \nu }\Box
-\nabla _{\mu }\nabla _{\nu }),  \label{frt2} \\
& f_{R}(R,T)=8\pi T_{\mu \nu }-f_{T}(R,T)T_{\mu \nu }-f_{T}(R,T)\Theta _{\mu
\nu }.
\end{align}%
In (\ref{frt2}), $R_{\mu \nu }$ is the Ricci tensor, $f_{R}(R,T)=\partial
f(R,T)/\partial R$, $f_{T}(R,T)=\partial f(R,T)/\partial T$, $\Box $ is the
D'Alambert operator, $\nabla _{\mu }$ is the covariant derivative and $%
\Theta _{\mu \nu }=-2T_{\mu \nu }-pg_{\mu \nu }$, with the energy-momentum
tensor $T_{\mu \nu }$ being considered the one of a perfect fluid.

Moreover, the covariant divergence of the energy-momentum tensor in $f(R,T)$ gravity reads \cite{alvarenga/2013,barrientos/2014}

\begin{eqnarray}\label{frt3}
\nabla^{\mu}T_{\mu\nu}&=&\frac{f_T(R,T)}{8\pi-f_T(R,T)}[(T_{\mu\nu}+\Theta_{\mu\nu})\nabla^{\mu}\ln f_T(R,T)\nonumber\\
&+&\nabla^{\mu}\Theta_{\mu\nu}-(1/2)g_{\mu\nu}\nabla^{\mu}T].
\end{eqnarray}

\section{The $f(R,T)=R+\protect\alpha R^{2}+\protect\gamma T$ model}

\label{sec:rar2lt}

\subsection{Field equations}

\label{ss:fes}

By substituting Eq.(\ref{i2}) in Eq.(\ref{frt2}) yields the following field
equations:

\begin{equation}  \label{fe2}
(2\alpha R+1)G_{\mu\nu}-\alpha R^{2}g_{\mu\nu}=(8\pi+\gamma)T_{\mu\nu}+\frac{%
\gamma}{2}(\rho-p)g_{\mu\nu}.
\end{equation}
In Eq.(\ref{fe2}), $G_{\mu\nu}$ is the usual Einstein tensor and we have
already taken the trace of the energy-momentum tensor of a perfect fluid to
be $\rho-3p$. The elegant form in which Eq.(\ref{fe2}) is presented makes
straightforward to recover GR when $\alpha,\gamma\rightarrow0$.

\subsection{Friedmann-like equations}

\label{ss:fle}

By defining the quantity

\begin{equation}
\Phi=\Phi(t)\equiv\left(\frac{\dot{a}}{a}\right)^{2}+\frac{\ddot{a}}{a},
\label{1}
\end{equation}
with $a=a(t)$ being the scale factor and dots representing time derivatives,
the non-null components of Eq.(\ref{fe2}) for a flat\footnote{%
In accordance with recent cosmic microwave background temperature
fluctuations observations \cite{hinshaw/2013}.} Friedmann-Robertson-Walker
metric are:

\begin{equation}
\Phi -\frac{\ddot{a}}{a}-12\alpha \Phi \left( \Phi -\frac{\ddot{a}}{a}+\Phi
^{2}\right) =\frac{1}{6}\left[ \left( 16\pi +3\gamma \right) \rho +\gamma p%
\right] ,  \label{fle1}
\end{equation}%
\begin{equation}
\Phi +\frac{\ddot{a}}{a}-12\alpha \Phi \left( \Phi +\frac{\ddot{a}}{a}-3\Phi
^{2}\right) =-\frac{1}{2}\left[ \left( 16\pi -3\gamma \right) p+\gamma \rho %
\right] .  \label{fle2}
\end{equation}

It is worthwhile reinforcing that, as required, the limits $\alpha ,\gamma
\rightarrow 0$ in Eqs.(\ref{fle1})-(\ref{fle2}) retrieve GR predictions.

Moreover, in thix context, Eq.(\ref{frt3}) is written as%
\begin{equation}
\dot{\rho}+3\frac{\dot{a}}{a}(\rho +p)=\tilde{\gamma}(\dot{\rho}-\dot{p}),
\label{w1}
\end{equation}

\noindent where $\tilde{\gamma}\equiv \bar{\gamma}/[2(1-2\bar{\gamma})]$ with $%
\bar{\gamma}\equiv \gamma /(\gamma -8\pi ).$ It is worth mentioning that by
making $\gamma \rightarrow 0$, GR is once again recovered.

\section{Analytical solutions for the scale factor and their cosmological
consequences}

\label{sec:assf}

As we can see in the previous section, the equations (\ref{fle1})-(%
\ref{fle2}) are nonlinear second-order differential equations. It is worth
pointing out that nowadays the nonlinearity is found in many areas of
Physics, including Condensed Matter \cite{1,1.1,1.2}, Field Theory \cite%
{2.1,2.2,2.3,2.4} and also Cosmology \cite{3,3.1,3.2}. In a cosmological
context, the nonlinear effects can play an important role to understand the
dynamics of the Universe. For instance, in a recent work it has been shown
that in a cosmological scenario with Lorentz symmetry breaking, the
so-called oscillons \cite{5} in the early Universe have passed through a
phase transition that changed their internal structure \cite{4}.

Unfortunately, as a consequence of the nonlinearity, in general we lose the
capability of getting the complete solutions. However, in this section we
will show that Equations (\ref{fle1})-(\ref{fle2}) can be solved
analytically in order to get the general solutions of the system.

In order to eliminate the explicit dependence on the term $\ddot{a}/a$, we
add the Eqs.(\ref{fle1}) and (\ref{fle2}) to conclude that

\begin{equation}
12\alpha\Phi^{2}(\Phi-1)+\Phi=\frac{8\pi}{3}\rho+\left( \frac {5\gamma}{3}%
-8\pi\right) p.  \label{4}
\end{equation}

It is important to remark that there is no restriction in adding these
equations and that such a mathematical approach was shown to be very useful 
\cite{6}. Also, in \cite{2.3}, it was used in order to find a class of
traveling solitons in Lorentz and \textit{CPT} breaking systems.

Now we will focus on getting analytical solutions for Eq.(\ref{4}). Looking
at it, it is natural to think that the functions $\rho $ and $p$ can be
represented by polynomial functions of third degree in $\Phi $. In fact,
such a representation is constantly used in studies concerning oscillon
theories \cite{7,7.1,7.2,7.3,7.4,m1.1}. In those cases, this mathematical
procedure allows to obtain the fundamental characteristics of the oscillons,
such as their field configuration, lifetime, amplitude and rate of decaying.
By using this approach we will have a specific class of solutions, but with
the great advantage of its analytical form.

Therefore, with the above motivation, we assume that $\rho $ and $p$ are
related by a general polytropic equation of state \cite{m1}:
\begin{equation}
p(t)=K\left[ \rho (t)\right] ^{\gamma _{0}}.  \label{5}
\end{equation}%
In (\ref{5}), $K$ and $\gamma _{0}$ are constants.

By substituting the above form of $p$ in Eq.(\ref{w1}), we obtain the
following constraints

\begin{equation}
\rho (t)=A_{0}[a(t)]^{-3/\Gamma _{0}},\text{ }p(t)=A_{0}K[a(t)]^{-3/\Gamma
_{0}},  \label{7}
\end{equation}%
where $A_{0}$ is an arbitrary constant of integration. Moreover, we are
using the following definition%
\begin{equation}
\Gamma _{0}\equiv \frac{(1-\tilde{\gamma})\left[ 1-\tilde{\gamma}K/\left( 1-%
\tilde{\gamma}\right) \right] }{1+K}.
\end{equation}
It is important to remark that, in order to avoid
singularities, we must impose that $\Gamma _{0}<0$.

On the other hand, by applying Eqs.(\ref{5}) and (\ref{7}) into (\ref{4}),
we find the equation

\begin{equation}
\Phi ^{3}+\Phi ^{2}-(1/12\alpha )\Phi =0,  \label{8}
\end{equation}%
\noindent where we are using the indentification 
\begin{equation}
K\equiv \frac{8\pi }{24\pi -5\gamma }.  \label{11}
\end{equation}



Now, in order to solve Eq.(\ref{8}) and consequently find a class
of analytical solutions for the scale factor, we impose that $\alpha =1/3$.
Thus, we can see from Eq.(\ref{8}) that there are two different roots for $%
\Phi $, which are given by 
\begin{eqnarray}
\Phi _{1} &=&0,  \label{q1} \\
\Phi _{2} &=&\frac{1}{2}.  \label{q2}
\end{eqnarray}

Thus, after some mathematical manipulations, we can obtain the following
analytical solutions for the scale factor

\begin{eqnarray}
a_{1}(t) &=&\sqrt{A_{1}t+B},  \label{19} \\
a_{2}(t) &=&A_{2}e^{-t/2}\sqrt{B_{2}e^{2t}+C}, \label{20}
\end{eqnarray}%
\noindent where $A_{i}$, $B_{i}$ and $C$ are arbitrary constants of integration, with $i=1,2$.

To interpret these solutions, we will construct the referred Hubble and deceleration parameters. The Hubble parameter, expressed by $H=\dot{a}/a$, shows us the expansion rate of the Universe in time, whereas the deceleration parameter, expressed by $q=-\ddot{a}a/\dot{a}^{2}$, is such that negative values stand for an accelerated expansion while positive values, for a decelerated expansion.

Let us start by  analysing solution (\ref{19}). Such a scale factor evolves in time according to Fig.\ref{fig1} below.

\begin{figure}[ht!]
\vspace{0.3cm} \centering
\includegraphics[height=5cm,angle=00]{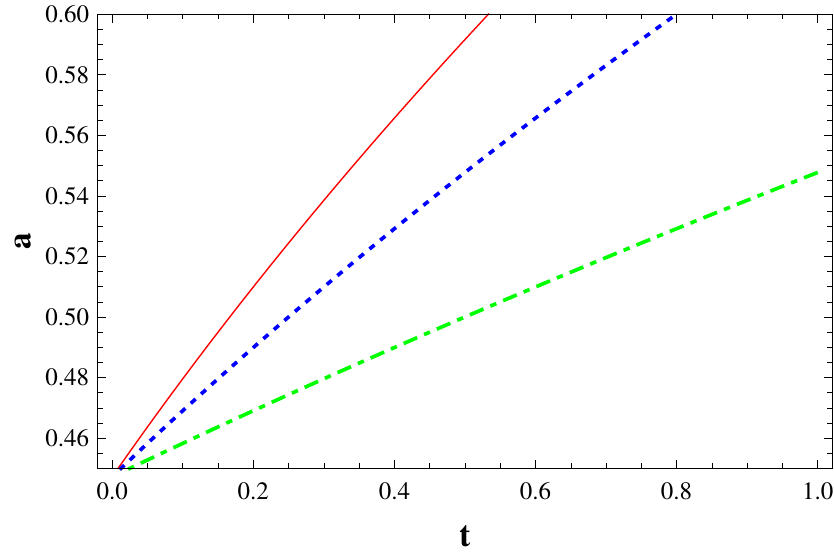}
\caption{Time evolution of the scale factor from Equation (%
\protect\ref{19}). The solid (red) line stands for $A_1=0.3$, while the dotted (blue) and dot-dashed (green) lines, for $A_1=0.2$ and $A_1=0.1$, respectively. For all curves, $B_1=0.2$.}
\label{fig1}
\end{figure}

\begin{figure}[ht!]
\vspace{0.3cm} \centering
\includegraphics[height=5cm,angle=00]{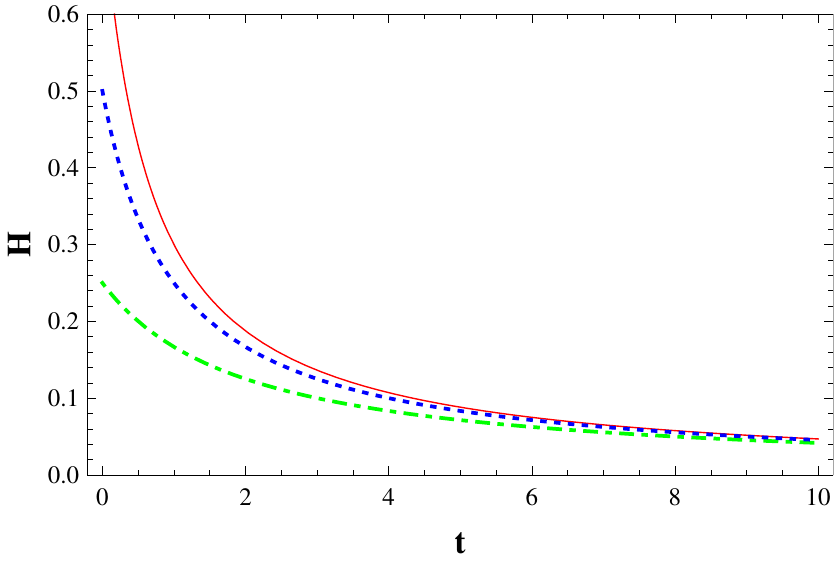}
\caption{Time evolution of the Hubble parameter from Equation (%
\protect\ref{21}). The solid (red) line stands for $A_1=0.3$, while the dotted (blue) and dot-dashed (green) lines, for $A_1=0.2$ and $A_1=0.1$, respectively. For all curves, $B_1=0.2$.}
\label{fig2}
\end{figure}

The referred Hubble parameter reads

\begin{equation} \label{21}
H_1=\frac{A_{1}}{2(A_{1}t+B_{1})},
\end{equation}
which is depicted in Fig.\ref{fig2}.

Moreover, independently of the values of the constants $A_{1}$ and $B_{1}$, Eq.(\ref{19}) yields $q_1=1$.

The behaviour of the cosmological parameters $a$, $H$ and $q$ obtained above
are in agreement with a universe dominated by radiation. In order to verify
this, let us recall that the standard Friedmann equations are obtained in
the present model by making $\alpha,\gamma=0$ in (\ref{fle1})-(\ref{fle2})
and read

\begin{equation}  \label{sfr1}
3\left(\frac{\dot{a}}{a}\right)^{2}=8\pi\rho,
\end{equation}
\begin{equation}  \label{sfr2}
2\frac{\ddot{a}}{a}+\left(\frac{\dot{a}}{a}\right)^{2}=-8\pi p.
\end{equation}

In order to make standard Friedmann equations above to describe a
radiation-dominated universe, one usually assumes $p=\rho/3$ as the EoS of
the Universe in (\ref{sfr2}). Such an assumption yields the solution $%
a(t)\sim t^{\frac{1}{2}}$, exactly as in Eq.(\ref{19}), obtained from the $%
f(R,T)$ formalism.

Furthermore, Fig.\ref{fig1} shows that $a\neq0$ as $t\rightarrow0$. In fact,
a null value for $a$ would indicate the origin of the Universe. However,
since we are treating the radiation dominated universe, $t=0$ does not
describe the Big-Bang. Rather, it describes the time in which radiation
starts dominating the Universe dynamics. In this way, the fact that $a\neq0$
for low values of time is in agreement with a radiation dominated universe.
We can also see that $a$ increases with time, corroborating an expanding
universe.

The Hubble parameter behaviour of Figure \ref{fig2} also strengthens our
argument. Firstly, we can see that it decreases with time, as it should happen in an expanding
universe. Secondly, since $H\sim t_H^{-1}$, with $t_H$ being the Hubble
time, at the end of the stage in which the Universe dynamics was dominated
by radiation, $H$ must be $\neq0$. High values of time in Fig.\ref{fig2}
(and also in Fig.\ref{fig1}) indicate the end of the radiation era rather
than the present or future epochs of the Universe, in which $H$
asymptotically tends to $0$. Such an asymptotically behaviour for $H$ can be
seen, for instance, in \cite{ms/2016,moraes/2014}, for which high values of
time stand for present and future epochs of the Universe evolution.

Moreover, the value which we obtained for the deceleration parameter, i.e., $%
q=1$, also is in accordance with a radiation dominated universe. The fact
that it is positive means that during this stage, the Universe expansion was
decelerating (in fact, the expansion started to accelerate some few billion
years ago \cite{hinshaw/2013}). Also, from the time proportionality obtained
for $a$ from the standard Friedmann equations above, i.e., $a\sim t^{\frac{1%
}{2}}$, the deceleration parameter definition $-\ddot{a}a/\dot{a}^{2}$
yields exactly $1$, i.e., our model has the same features of a standard
cosmology radiation-dominated universe.

Now, using Eq.(\ref{20}), we find the following results for the cosmological parameters

\begin{equation} \label{22}
H_2=\frac{A_{2}e^{-\frac{t}{2}}(B_{2}e^{2t}+C)}{2\sqrt{C-B_{2}e^{2t}}t},
\end{equation}
\begin{equation} \label{23}
q_2=\frac{C(6B_{2}e^{2t}-1)+B_{2}^{2}e^{4t}}{(C+B_{2}e^{2t})^{2}}.
\end{equation}

The evolution of these quantities in time can be appreciated in Figs.\ref{fig4}-\ref{fig5} below.


\begin{figure}[ht!]
\vspace{0.3cm} \centering
\includegraphics[height=5cm,angle=00]{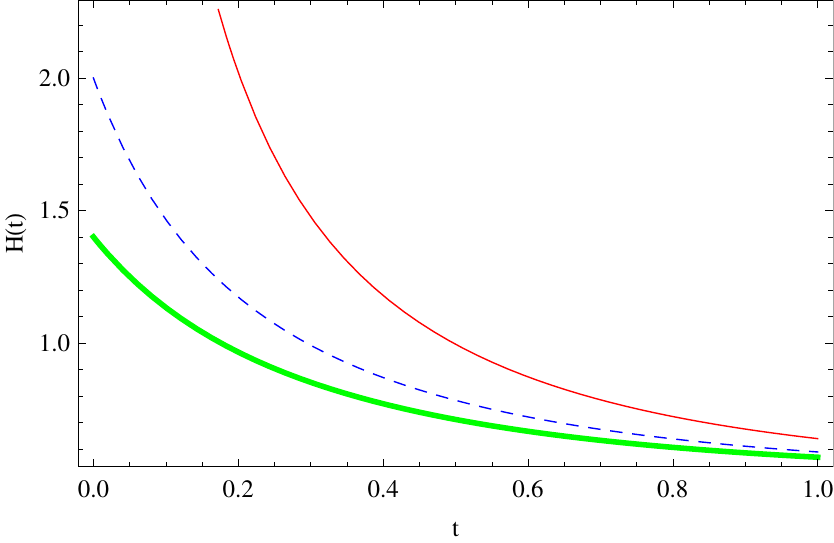}
\caption{Time evolution of the Hubble parameter from Equation 
(\protect\ref{22}). The (blue) dotted line stands for $A_{2}=2$ and $B_{2}=1.5$, the (green) dot-dashed stands for $A_{2}=3$ and $B_{2}=1.9$ and (red) solid lines stand for $A_{2}=B_{2}=1$. In all
curves, $C=-0.9$.}
\label{fig4}
\end{figure}

\begin{figure}[ht!]
\vspace{0.3cm} \centering
\includegraphics[height=5cm,angle=00]{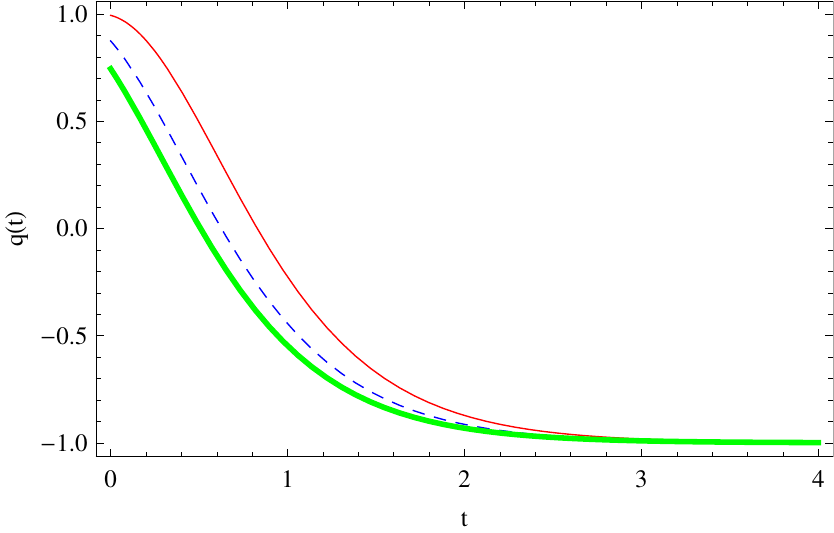}
\caption{Time evolution of the deceleration parameter from Equation (\protect\ref{23}%
). The (blue) dotted line stands for $A_{2}=2$ and $B_{2}=1.5$, the (green) dot-dashed stands for $A_{2}=3$ and $B_{2}=1.9$ and (red) solid lines stand for $A_{2}=B_{2}=1$. In all
curves, $C=-0.9$.}
\label{fig5}
\end{figure}

The cosmological model constructed from Eq.(\ref{20}) is quite more general than the one presented earlier. This feature can easily be checked by investigating Fig.\ref{fig5}. For different values of the constants involved, $q_2$ departs from $1$, which stands for a radiation-dominated era. As time passes by, $q$ assumes the value $0.5$, which is the deceleration parameter of a matter-dominated universe. This can be checked by taking $p=0$ in Eq.(\ref{sfr2}). Fig.\ref{fig5} also shows that the model predicts a transition from a decelerated to an accelerated phase of expansion of the Universe, since the deceleration parameter eventually assumes negative values. These values are in agreement with observations, as one can check, for instance, the 192 ESSENCE SNe Ia data \cite{lu/2008}. 

\section{Discussion}

\label{sec:dis}


It is known that for a small but non-negligible period of time the dynamics of the early universe
was dominated by radiation. During this epoch, the density and temperature
of photons were high enough to prevent atoms, (and consequently) stars and
galaxies to form.

In such a stage, the EoS of the Universe is written as $p=\rho/3$. For a
perfect fluid, such an EoS yields a null trace of the energy-momentum tensor
and therefore one expects, in this regime, $f(R,T)$ gravity to simply
retrieve $f(R)$ gravity. Indeed, no contributions from the former are
expected since the dependence on $T$ disappears.

Such an $f(R,T)$ formalism shortcoming has generated some important
discussions. In \cite{ms/2016}, in order to surpass such an unpleasant
feature, the authors have formulated a cosmological scenario for the $%
f(R,T^{\phi})$ gravity, with $\phi$ being a scalar field. They have showed
that even in the regime $T=0$, the field equations of the model present
extra contributions, when compared to those from $f(R)$ gravity, coming from
the trace of the energy-momentum tensor of the scalar field. Such a
formalism originated the possibility of studying gravitational waves in $%
f(R,T)$ gravity \cite{amam/2016} (recall that the $T=0$ regime is also
obtained in vacuum).

Here, instead, we have proposed a quadratic correction for the $R$%
-dependence of the $f(R,T)$ function. Motivated by the application of the SM
in $f(R)$ cosmology \cite%
{starobinsky/2007,goswami/2013,nojiri/2015,abebe/2016} and $f(R,T)$
astrophysics \cite{zubair/2015,noureen/2015,noureen/2015b}, we intended here
to check if from the $f(R,T)=R+\alpha R^{2}+\gamma T$ theory, one could
derive a healthy cosmological scenario.

In constructing our model, we have obtained a highly nonlinear set of
differential equations for the scale factor $a$, from which important
and informative cosmological parameters are obtained. 

Remarkably, for small values of time, the values of our scale factor
solution presented in Fig.\ref{fig1} are not close to $0$. The restriction of this model
to the radiation era of the Universe can be checked also in Fig.\ref{fig2},
in which we can see that for high values of time (end of radiation era) the
Hubble parameter does not tends asymptotically to $0$, which is expected in
a recent universe (check, for instance, \cite{ms/2016}).

We have presented from solution (\ref{19}) a formalism which makes $f(R,T)$ gravity able to
generate a cosmological scenario in which radiation dominates the dynamics
of the Universe. The relevance of such a construction lies on the fact that
one does not expect $f(R,T)$ gravity to be capable of describing such a
stage of the Universe without simply recovering $f(R)$ gravity. Here,
instead, we have shown that besides predicting a variety of well behaved
cosmological and astrophysical scenarios in $f(R)$ gravity, the SM within
the $f(R,T)$ gravity solves the $T=0$ issue of $f(R,T)$ theories.

On the other hand, solution (\ref{20}) is related to a more complete cosmological scenario. It predicts, from the analysis of the referred deceleration parameter, the radiation, matter and dark energy-dominated eras, as well as the transition among these stages, in a continuous form, which is certainly a milestone in theoretical cosmology.

\begin{acknowledgements}
PHRSM would like to thank S\~ao Paulo Research Foundation (FAPESP), grant 2015/08476-0, for financial support. RACC thanks to  CAPES for financial support.
\end{acknowledgements}



\end{document}